\documentstyle[prb,aps,multicol,epsf]{revtex}
\begin{document}
\draft
\title{Regimes of correlated hopping via a two-site
interacting chain}
\author{A.~D.~Ballard and M.~E.~Raikh}
\address{
 Department of Physics, University of Utah, Salt Lake City,
Utah 84112, U.S.A.}
\date{\today}
\maketitle
\tighten
\begin{abstract}
Inelastic  transport of electrons through a two-impurity chain
is studied theoretically with account of intersite Coulomb
interaction, $U$. Both limits of ohmic transport (at low bias) and strongly 
non-ohmic transport (at high bias) are considered. 
We demonstrate that correlations, induced by a finite $U$,
in conjunction with  conventional Hubbard correlations, 
give rise to a distinct transport regime, with current 
governed by {\em two-electron}
hops. This regime realizes when a single-electron hop onto the 
chain and a single-electron hop      
out of the chain are both ``blocked'' due to the finite $U$,
so that  conventional correlated 
sequential 
transport is impossible. 
The regime of two-electron hops manifests itself in the form of an additional step  in
the current-voltage characteristics, $I(V)$.
\end{abstract}


\begin{multicols}{2}
  
\section{INTRODUCTION}

A model of electron transport via two localized 
states positioned in sequence has been studied theoretically
for more than two decades\cite{levin82,larkin87,glazman88,chase89,raikh92,bahlouli94,kinkhabwala00,raikh02}.
There are several reasons why this model attracts the attention
of researchers: ({\em i}) the model is tractable analytically;
({\em ii}) it yields nontrivial predictions that are in 
agreement with
experimental observations\cite{xu90,ephron92,ephron94,xu95,savchenko95,yoshida97,dai01}; 
({\em iii})
it captures physics that is much richer than tunnel transport
through a single site. In particular, the two-site model 
allows one to reveal various aspects of {\em correlated} 
transport, 
that are  relevant for the bulk systems, and to treat 
correlated transport {\em exactly}.

In general, for the bulk hopping conductivity,
two mechanisms are known to cause  
correlation in the time moments of
different hops, thus giving rise
to the correlated character of transport.


The first mechanism is the {\em on-site}
Hubbard repulsion, $U_0$. In the limit  $U_0\rightarrow \infty$,
the occupancy of a given site is restricted
to the values of $0$ or $1$. The resulting
correlation (dubbed as Hubbard correlation\cite{book84})
reduces to  the requirement that electron hops 
occur only between singly occupied and empty sites.

Calculation of the current through a two-site chain
with Hubbard correlations was
carried out in 
Refs.\onlinecite{levin82,chase89,raikh92,kinkhabwala00}
both for ohmic\cite{levin82,chase89} 
and strongly non-ohmic\cite{raikh92,kinkhabwala00} regimes.
These calculations, being exact, allow one to test the
applicability of a standard mean-field description\cite{book84}
of the hopping transport.  Within this description, the time
moments of  different hops are assumed completely
uncorrelated, while the probability of a given hop is 
determined by {\em average} occupations of the 
constituting sites. 

For finite  $U_0$, double occupation of a single site
becomes possible. Together with suppression of the
Hubbard correlations, finite $U_0$ makes the transport
through a chain spin-dependent\cite{bahlouli94,ephron94}.
This is because the second electron hopping on a given 
singly occupied site must, by virtue of the Pauli
principle,  have the opposite spin.      

The second mechanism, that causes  correlation 
between  different hops
is the {\em intersite}
Coulomb interaction, $U$. Due two a finite $U$,
the energy position of a given site, and, thus,
the probability of an electron hop onto this site, 
depends on the occupation of the neighboring sites.   
The energy $U$ can be viewed as an additional 
{\em charging energy} required for an electron
to hop onto and out of the two-site chain. 

On average, such a charging, being analogous
to the Coulomb blockade, impedes the hopping transport. 
For bulk systems this mechanism is commonly
accounted for within the mean-field theory\cite{book84}.  
The key assumption adopted to 
incorporate charging into the standard 
scheme of calculation of the hopping 
conductivity is that this charging amounts 
exclusively to the depletion of the density 
of states near the Fermi level\cite{efros75}. 
Under this assumption,  strongly correlated
time evolution of the populations
of {\em all} sites is replaced by  completely
uncorrelated evolution of populations  
of {\em much fewer} sites, the argument used 
for such a replacement being that
only the hops between these fewer sites
govern the transport.

Another effect of the intersite Coulomb 
interaction, which had received much less attention, 
is that this interaction opens the possibility for 
{\em many-electron} hops\cite{knotek73}. In the course of
such hops, two (or more) electrons change
their spatial positions upon absorption (or emission)
of a {\em single} phonon.  
This process is analogous to the light absorption
in the helium atom\cite{helium}, 
in course of which two electrons
can be excited by one photon. 
Obviously, such an absorption is
possible only due to 
the electron-electron interactions.
It is also apparent that, as a new transport channel, 
many-electron hops {\em facilitate} the transport.

In the early paper Ref.\onlinecite{knotek73},
where many-electron hops were first treated 
analytically, the roles of the two electrons,
participating in the two-electron transition,
were very dissimilar. While the transfer of the
first electron occurred between the sites
belonging to the current-carrying network,
the initial and final states of the second
electron did not belong to the network.
Consequently,  facilitation of transport
by two-electron transitions amounted to the
effective reduction of the activation energies 
of single-electron hops within the  
current-carrying network.

Later, on the basis of numerical simulations
\cite{tenelsen95,perez97} it was concluded that 
many-electron hops may constitute a significant
portion of the current-carrying path through
the sample. However, no analytical theory
of hopping transport with account of many-electron 
transitions has been developed so far. 
This is because all existing theories are 
based  on introducing  effective resistors 
between the pairs of sites, which is
impossible in the presence of 
many-electron transitions.

In the present paper, we take advantage 
of the fact that the two-site model is 
exactly solvable and thus allows one to study
the interplay of 
all three correlation effects, namely, 
Hubbard correlations, Coulomb-blockade-induced 
correlations, and correlated two-electron hops.
The fact that the first two correlation mechanisms
impede the transport, while the third one facilitates 
it, suggests that such an interplay is nontrivial. 
 
Our main result is a demonstration of a distinct
regime of transport through the two-site chain,
in which  two-electron hops {\em dominate} the 
passage of current. This regime becomes
possible due to intersite interaction, $U$,
when both manifestations of this interaction
are at work: ({\em i}) finite $U$ allows two-electron
transitions, and ({\em ii}) it blocks both individual 
single-electron transitions onto and out of the chain.
Conceptually, this interaction-dominated transport regime is analogous
to inelastic cotunneling through a Coulomb blockaded
quantum dot\cite{averin90,glazman90,review}. 
However, in terms of its manifestations, the 
important difference between the dot and the
two-site chain is that the dot contains many levels,
so that the cotunneling rate is a sum of many contributions.
By contrast, the transport through a blocked two-site chain
is governed by a single two-electron hop, leading to 
distinctive temperature and bias dependencies of current. 

We first identify the regime of two-electron hops for strongly 
non-ohmic transport at zero temperature, where it
manifests itself in the form of additional steps in
the current-voltage characteristics, $I(V)$. 
Then we demonstrate the relevance of two-electron
hops for the ohmic transport.
\section{HOPPING TIMES}
The two-site model is illustrated in Fig. 1.
Under the applied bias, $V$, the Fermi levels 
in the left and right leads are shifted $V/2$
and $-V/2$, respectively. The two sites, $1$ and $2$, 
are located at distances $d_1$ and $d_2$ with respect
to the center of the barrier of thickness, $D$.
We adopt the definition of the energy position
of site $1$ to be $\varepsilon_1$, when site $2$
is {\em empty}. Consequently, when site $2$ is
occupied, the energy position of  site $1$ is
$ \varepsilon_1+U$. Analogously, $\varepsilon_2$ and
$\varepsilon_2+U$ are the energy positions of 
site $2$ for empty and occupied site $1$, respectively.
We assume that there is tunnel coupling with amplitude,
$t_{1,k}$,  between site $1$ and the extended state, {\bf k},
in the left lead.  Similarly, the state, {\bf p}, in the
right lead is coupled to site $2$ with amplitude, $t_{2,p}$.
Then the waiting time, $\tau_{\mbox{\tiny 1}}$, for an electron 
in the left lead to
tunnel onto  site $1$ is given by 
$\tau_{\mbox{\tiny 1}}^{-1}=f_l(\varepsilon_1)\Gamma_l(\varepsilon_1)/\hbar$,
where the tunneling width, $\Gamma_l(\varepsilon)$, is defined as
\begin{equation}
\label{width1}
\Gamma_l(\varepsilon)=2\pi
\sum_{\mbox{\tiny\bf k}}
\vert t_{1,k}\vert^2\delta (\varepsilon-
\varepsilon_{\mbox{\tiny\bf k}}) \propto
\exp\left[-\frac{D-2d_1}{a}\right].
\end{equation}
The corresponding expression for the waiting time, 
$\tau_{\mbox{\tiny 3}}$, 
for electron on  site $2$
to escape into the right lead reads 
$\tau_{\mbox{\tiny 3}}^{-1}=
\Bigl[1- f_l(\varepsilon_2)\Bigr]\Gamma_r(\varepsilon_2)/\hbar$,
 with
\begin{equation}
\label{width2}
\Gamma_r(\varepsilon)=2\pi
\sum_{\mbox{\tiny\bf p}}
\vert t_{2,p}\vert^2\delta (\varepsilon-
\varepsilon_{\mbox{\tiny\bf p}}) \propto
\exp\left[-\frac{D-2d_2}{a}\right].
\end{equation}
Here $a$ is the localization
radius of the on-site wave functions;
$f_l$, $f_r$ are the Fermi distribution functions
in the left and right leads, respectively, which determine the 
temperature dependencies of $ \tau_{\mbox{\tiny 1}}$ and
$ \tau_{\mbox{\tiny 3}}$. 

Transition $1\rightarrow 2$ involves tunneling accompanied by the 
emission of a phonon with energy $\varepsilon_1-\varepsilon_2$. Thus,
the temperature dependence of the corresponding time,
$ \tau_{\mbox{\tiny 2}} \propto \exp[2(d_1+d_2)/a]$, is weak.
\begin{figure}
\narrowtext
{\epsfxsize=8.5cm
\centerline{\epsfbox{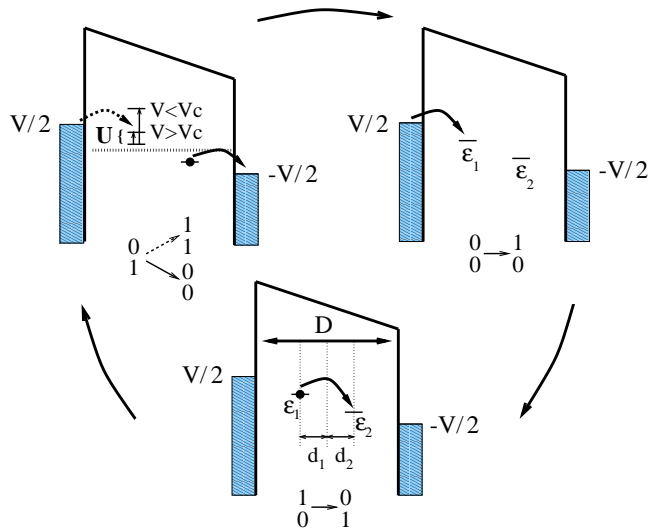}}}
\protect\vspace{0.5cm}
\caption{Schematic illustration of the two-site model.  Evolution 
of the occupation numbers of sites in the sequential hopping regime
for blocked ($V<V_c$) and unblocked ($V>V_c$) transition 
$l\rightarrow 1$.} 
\label{fig1}
\end{figure}

In all previous analytical treatments
\cite{levin82,larkin87,glazman88,chase89,raikh92,bahlouli94,kinkhabwala00,raikh02} 
of the two-site
model it was assumed that the passage of current is 
governed by only three transitions: $l\rightarrow 1$, $1\rightarrow 2$,
and $2\rightarrow r$. Our key point in this study is that the
tunnel couplings $t_{1,k}$  and  $t_{2,p}$ together with finite
intersite interaction, $U$, allow for the additional transition,
$(l,2)\rightarrow (1,r)$, that involves simultaneous
change of occupation of {\em both} sites. Moreover, we will
demonstrate that, within a certain bias range, these two-electron
transitions dominate the passage of current through the two-site
chain.   Calculation of the waiting time, $\tau_{\mbox{\tiny c}}$, 
for the  two-electron transition is similar to the calculation
\cite{review}
of the inelastic cotunneling rate through a quantum dot
\begin{eqnarray}
\label{width12}
\frac{1}{\tau_{\mbox{\tiny c}}}\!=\frac{2\pi}{\hbar}
\sum_{\mbox{\tiny\bf k}, \mbox{\tiny\bf p}}
\vert t_{1,k}t_{2,p}\vert^2\Biggl
[\frac{1}{\varepsilon_2-\varepsilon_{\mbox{\tiny\bf p}}}+
\frac{1}{\varepsilon_{\mbox{\tiny\bf k}}-
\varepsilon_1-U}\Biggr]^2 
\times
\cr
 \delta\left(\varepsilon_{\mbox{\tiny\bf k}}+
\varepsilon_2-\varepsilon_{\mbox{\tiny\bf p}}-\varepsilon_1\right)
f_l\left(\varepsilon_{\mbox{\tiny\bf k}}\right)
\Bigl[1-f_r(\varepsilon_{\mbox{\tiny\bf p}})\Bigr],
\end{eqnarray} 
where the two terms in the square brackets originate from 
the intermediate states with empty and doubly-occupied chain, 
respectively. The many-body nature  of the two-electron transition
manifests itself in the fact that these terms cancel each other
in the limit $U\rightarrow 0$. 

With the help of  definitions (\ref{width1}) and (\ref{width2}),
Eq. (\ref{width12}) can be presented as

\vspace{3mm}

\begin{equation}
\label{form}
\frac{1}{\tau_{\mbox{\tiny c}}}\!=\!U^2\!\!\int\! d\varepsilon 
\frac{\Gamma_l(\varepsilon_1-\varepsilon)\Gamma_r(\varepsilon_2-\varepsilon)}
{2\pi\hbar~\varepsilon^2(\varepsilon +U)^2}f_l(\varepsilon_1-\varepsilon)
\Bigl[1-f_r(\varepsilon_2-\varepsilon)\Bigr].
\end{equation}

\vspace{3mm}

In principle, the
integrand needs to be regularized at the resonances, $\varepsilon = 0, -U$,
following, {\em e.g.},  the procedure of Ref. \onlinecite{Turek02}. 
However, the corresponding singular contributions are
 negligibly small in the cases of interest,  $|\varepsilon_1|,
 |\varepsilon_2| \gg T$, where $T$ is the temperature.

We will establish the analytical form of $\tau_{\mbox{\tiny c}}$ in the limit
of large enough bias, $V$, when it is temperature-independent.
In this limit, the integration in Eq. (\ref{form}) is restricted to
the interval $\varepsilon_1-V/2 < \varepsilon < \varepsilon_2+V/2$,
which is set by the  distribution functions in the leads.
Neglecting the weak energy dependencies of $\Gamma_l$ and $\Gamma_r$
and evaluating the integral, we obtain
\begin{equation}
\label{evaluate}
\frac{1}{\tau_{\mbox{\tiny c}}}=\frac{\Gamma_l\Gamma_r}{\pi\hbar~U}
\Biggl[\mbox{\large$\Phi$}\left(\frac{2\varepsilon_1-V}{2U}\right)-
\mbox{\large$\Phi$}\left(\frac{2\varepsilon_2+V}{2U}\right)\Biggr],
\end{equation}
where the function $\mbox{\large$\Phi$}(x)$ is defined as
\begin{equation}
\label{phi}
\mbox{\large$\Phi$}(x)=\frac{2x+1}{2x(x+1)}-\ln\left(\frac{x+1}{x}\right).
\end{equation}
The  asymptotes
of $\tau_{\mbox{\tiny c}}$ in the limits of small and large intersite
interaction, $U$, can easily be found from \newline Eqs. (\ref{evaluate}), (\ref{phi})
\begin{equation}
\label{smallU}
\frac{1}{\tau_{\mbox{\tiny c}}}=\frac{4\Gamma_l\Gamma_r U^2}{3\pi\hbar}
\Biggl[\frac{1}{\left(2\varepsilon_1-V\right)^3}-\frac{1}{\left(2\varepsilon_2+V\right)^3}\Biggr], 
~U\rightarrow\!0,
\end{equation}


 
\begin{equation}
\label{largeU}
\frac{1}{\tau_{\mbox{\tiny c}}}=\Biggl(\frac{2\Gamma_l\Gamma_r}
{\pi\hbar }\Biggr)\frac{\varepsilon_2-\varepsilon_1+V}{(2\varepsilon_1-V)(2\varepsilon_2+V)}~, 
~U\rightarrow \infty.
\end{equation} 
It is worth noting that the rate Eq. (\ref{width12}) of the 
interaction-induced hop becomes $U$-independent in the
limit of strong intersite interaction, as follows from Eq. (\ref{width12})
as well as from Eq. (\ref{largeU}).
Note also that in both limits, the rate $ \tau_{\mbox{\tiny c}}^{-1}$ is proportional
to $(\varepsilon_2-\varepsilon_1+V)$, which plays the role of ``phase volume'' for
two-electron transitions, as we will see below.

It is seen from Eq. (\ref{width12}) that 
$\tau_{\mbox{\tiny c}} \propto \exp~[2(D-d_1-d_2)/a]$,
i.e. it is shorter than the waiting time for direct tunneling,
which is $\propto \exp~(2D/a)$. Important, however, is that 
$\tau_{\mbox{\tiny c}}$ can be comparable to $\tau_{\mbox{\tiny 2}}$.
As we will see later, transport, dominated by two-electron 
transitions, is most prominent when $\tau_{\mbox{\tiny c}}$ and
$\tau_{\mbox{\tiny 2}}$ are of the same order.

\section{NON-OHMIC REGIME}
\subsection{QUALITATIVE DISCUSSION}
In this Section we assume that the temperature
is zero, so that hopping transport through
a chain is possible if all hops, involved in
the passage of current, are {\em activationless}.
All throughout the paper we assume the on-site
Hubbard repulsion, $U_0$, to be infinite. 
To classify different transport regimes, we
first note that when the bias, $V$, is big enough, 
$V \gg U$, the intersite repulsion, $U$, can be
neglected. The current path then consists of three 
hops, $l\rightarrow 1$, $1\rightarrow 2$, and $2\rightarrow r$,
in {\em arbitrary} order, in the sense that the only condition
for a hop to occur is that the initial state is occupied, while the
final state is empty. In particular, as illustrated in Fig. 1,
the waiting time for the hop $l\rightarrow 1$ does not depend on whether
 site $2$ is occupied or empty. Similarly, the waiting time for
the hop $2\rightarrow r$ is independent of the occupation of  site $1$.
 This, however, does not mean that the transport in the regime
of high bias is completely uncorrelated. It only means that the 
correlations are of purely Hubbard origin, i.e. the electron in the
left lead has to ``wait'' for the hop $1\rightarrow 2$, leaving site
$1$ empty, to occur.  This Hubbard-correlated
regime takes place for $V>V_c$, where $V_c$ is determined by the
condition $\varepsilon_1+U=V_c/2$. As seen from Fig. 1, 
at $V=V_c$,  site $1$, shifted upward by $U$, due
to site $2$ being occupied, becomes aligned with the Fermi level
in the left lead.  
  
As the bias is reduced below $V_c$, the intersite repulsion changes
radically the electron dynamics. Now the activationless hop $l\rightarrow 1$
is possible {\em only} if site $2$ is empty (see Fig. 1). When site 2
is occupied, the transition $l\rightarrow 1$ is ``blocked''.
Clearly, 
the average current drops down in a step-like fashion as the  bias is swept through $V_c$.
It is also clear that below $V_c$ the current is more correlated than above $V_c$.
This is because the hops $l\rightarrow 1$, $1 \rightarrow 2$, and
$2\rightarrow  r$ occur in a strict succession.

The two above regimes were discussed in the literature before\cite{raikh92,kinkhabwala00}.
However, the  most nontrivial scenario of activationless passage of current
unfolds when {\em both} transitions  $l \rightarrow 1$
and $2\rightarrow r$ are blocked. The concept of inelastic cotunneling 
through a dot\cite{averin90,glazman90,review}
suggests that in this case the activationless current is due to the
two-electron transition $(l,2) \rightarrow (1,r)$. However, the theory of inelastic
cotunneling implies that the two-electron transition 
is {\em immediately} followed by {\em relaxation}. For a two-site chain this relaxation is 
a single-electron transition $1\rightarrow 2$. Then the ``immediate'' relaxation 
requires that sites $1$ and $2$ are spatially close to each other (as is the case for dots).
Note, however, that the spatial proximity of sites $1$ and $2$, meaning that $d_1, d_2 \ll D$,
makes the two-electron time $\tau_c \propto \exp[2(D-d_1-d_2)/a]$ \{see Eq.~(\ref{width12})\} 
quite long, and actually comparable to the direct-tunneling time. Thus, in order to
yield a significant current, the separation, $(d_1+d_2)$, should be $\sim D$. On the other
hand, large separation of sites $1$ and $2$ unavoidably opens up new relaxation channels, which
are ``reverse'' single-electron transitions $1\rightarrow l$ and $r\rightarrow 2$. Obviously,
these transitions ``undo'' the two-electron transition and forbid the current to flow.
Then, it is quite nontrivial that for certain  energy \
configurations $\{\varepsilon_1,\varepsilon_2\}$ 
of the sites, the reverse relaxation channels are blocked, leaving
the transition $1\rightarrow 2$, 
the {\em only} allowed transition that can follow a two-electron hop. 
At the same time, the transition $1\rightarrow 2$, happening after
the two-electron hop, {\em completes the current cycle}.
We will dub such configurations with unidirectional relaxation as ratchet
configurations.

\subsection{RATCHET-TYPE CONFIGURATIONS}

Consider the configuration of sites depicted in Fig. 2.
Site $2$ is lower than $-V/2$ and thus is occupied. The 
energy, $\varepsilon_1 + U$, of site $1$ is above $V/2$,
so that site $1$ is empty. For this configuration both
transitions $l\rightarrow 1$ and $2\rightarrow r$ are 
blocked. The condition that the two-electron transition 
$(l,2) \rightarrow (1,r)$ is allowed at zero temperature
reads

\begin{equation}
\label{condition1}
\frac{V}{2}-\left(-\frac{V}{2}\right)=V>\left(\varepsilon_1-\varepsilon_2\right).
\end{equation}
The lhs in Eq. (\ref{condition1}) is the minimal energy required to transfer an electron 
between the leads, while the rhs is the energy required to transfer an electron between
sites $2$ and $1$. 
\begin{figure}
\narrowtext
{\epsfxsize=8.5cm
\centerline{\epsfbox{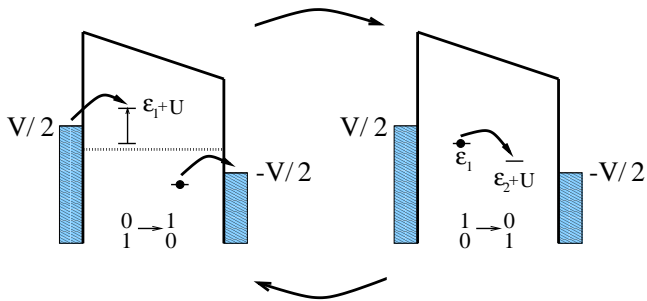}}}
\protect\vspace{0.5cm}
\caption{Schematic illustration of the non-ohmic transport
regime dominated by two-electron hops $(l,2) \rightarrow (1,r)$
followed by an inelastic transition $1\rightarrow 2$.
For the ratchet-type configuration shown, both ``backwards''
transitions following  $(l,2) \rightarrow (1,r)$ are forbidden.} 
\label{fig5}
\end{figure}
The condition allowing a two-electron transition has the form
Eq. (\ref{condition1}) since this transition can be viewed as a transfer of an electron
from the left to the right lead, accompanied by a ``backward'' excitation $2\rightarrow 1$ of the
other electron. Obviously, the ``direct'' transition $1\rightarrow 2$ can occur 
after the two-electron hop. The ratchet-type configuration is such that this direct transition
is not preceeded by other single-electron transitions. This would absolutely  be the case 
if the following two conditions are met

\vspace{5mm}

\begin{eqnarray}
\label{condition2}
\varepsilon_1 &<& \frac{V}{2}, \\
\label{condition3}
\varepsilon_2+U &>& -\frac{V}{2}.
\end{eqnarray}


Indeed, the first condition ensures that the electron from the occupied site $1$ cannot
hop back into the left lead, while the second condition guarantees that the empty site
$2$ does not get occupied as a result of an electron hop from the right lead. 
Our prime observation is that ratchet-type configurations are possible, i.e.
all five conditions, $(\varepsilon_1 +U) > V/2$, $\varepsilon_2 < -V/2$, and
Eqs. (\ref{condition1})-(\ref{condition3}), are satisfied simultaneously
 within a {\em finite} domain on the $(\varepsilon_1,\varepsilon_2)$ plane. 
 This domain corresponds to the triangular region in Fig. 3, where $I_c$ denotes
the magnitude of current in the regime of two-electron hops. This region
is situated adjacently to two rectangular domains, within which one of
the transitions $l\rightarrow 1$ or $2\rightarrow r$ is blocked. $I_1$ stands for
the magnitude of the strictly sequential current in these domains. Finally, within
the  domain $\left(~\varepsilon_1 < -U+V/2,~\varepsilon_2 > -V/2~\right)$ both transitions are
unblocked, so that the current, $I_2$, is limited only by the Hubbard correlations.  
\begin{figure}
\narrowtext
{\epsfxsize=8.5cm
\centerline{\epsfbox{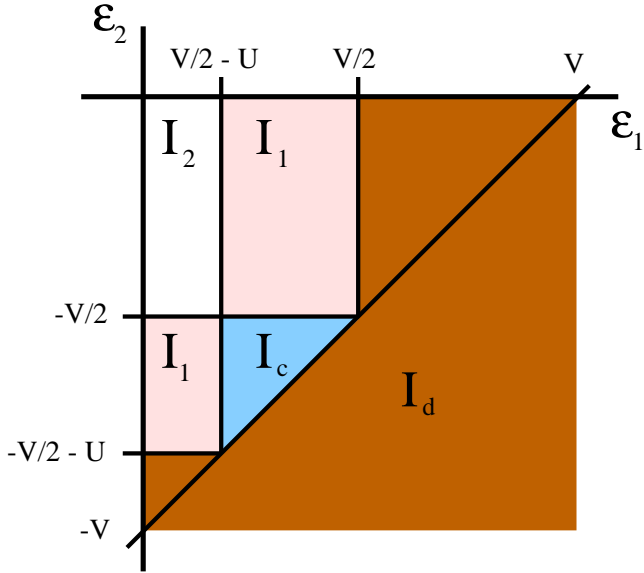}}}
\protect\vspace{0.5cm}
\caption{(Color online)  Domains of the site energies 
corresponding to the different regimes of transport.} 
\label{fig4}
\end{figure}
To summarize our qualitative analysis, in Fig. 4 we depict schematically the current-voltage
characteristics at zero temperature.
In the absence of interactions the $I-V$ curve would exhibit only one 
step, from the ``long-hop'' current, $I_d$, corresponding to direct tunneling
between the leads, to Hubbard-correlated current,
$I_2$. Interactions give rise to two additional steps: from $I_d$ to 
``doubly-blockaded'' current, $I_c$, and from $I_c$ to ``singly-blockaded'' 
current, $I_1$. 
\begin{figure}
\narrowtext
{\epsfxsize=8.5cm
\centerline{\epsfbox{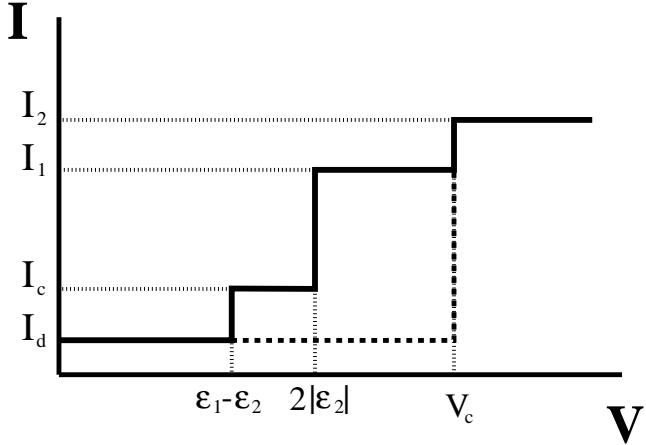}}}
\protect\vspace{0.5cm}
\caption{Schematic $I-V$ characteristics of the two-site
chain at zero temperature is shown for the case 
$-U< \varepsilon_1+\varepsilon_2 < 0$.  
Weak bias dependencies of current within
each step are neglected.} 
\label{fig3}
\end{figure}
In the remainder of this Section we calculate the values
$I_1$, $I_2$, and $I_c$, and study the correlation 
characteristics of transport within each step.

\subsection{CALCULATION OF CURRENT}

For infinite Hubbard repulsion,  possible sets of
the occupation numbers $(n_1, n_2)$  of the sites $1$ and $2$
are restricted to $(0,0)$, $(0,1)$, $(1,0)$, and $(1,1)$. 
Following Refs.~\onlinecite{levin82,chase89}
we introduce the probabilities, $ {\huge P}_{n_1,n_2}$, of each
set. Normalization requires that 
$ {\huge P}_{0,0}+ {\huge P}_{0,1}+ {\huge P}_{1,0}+ {\huge P}_{1,1}=1$.   

As discussed above, in the regime $V>V_c$, when both transitions $l\rightarrow 1$
and $2\rightarrow r$ are unblocked, the transport involves all four sets
of the occupation numbers. For $V<V_c$, when $l\rightarrow 1$ is blocked,
double occupation is prohibited, i.e. $ {\huge P}_{1,1}=0$. The master 
equations for probabilities $ {\huge P}_{n_1,n_2}$ can be cast in a
form that accounts for both cases, i.e. applies within the entire interval
$-V/2 < \varepsilon_2 < \varepsilon_1 < V/2$,  as follows
\begin{eqnarray}
\label{derivP10}
\frac{d\!~{\huge P}_{1,0}}{d\!~t}&=&
-\frac{
{\huge P}_{1,0}}
{\tau_{\mbox{\tiny 2}}}+
\frac{
{\huge P}_{0,0}}
{\tau_{\mbox{\tiny 1}}}+
\frac
{{\huge P}_{1,1}}
{\tau_{\mbox{\tiny 3}}}\Theta(V-V_c),\\
\label{derivP01}
\frac{d\!~{\huge P}_{0,1}}{d\!~t}&=&
-\frac{
{\huge P}_{0,1}}
{\tau_{\mbox{\tiny 3}}}-
\frac{
{\huge P}_{0,1}}
{\tau_{\mbox{\tiny 1}}}\Theta(V-V_c)+
\frac
{{\huge P}_{1,0}}
{\tau_{\mbox{\tiny 2}}},\\
\label{derivP00}
\frac{d\!~{\huge P}_{0,0}}{d\!~t}&=&
-\frac{
{\huge P}_{0,0}}
{\tau_{\mbox{\tiny 1}}}+
\frac{
{\huge P}_{0,1}}
{\tau_{\mbox{\tiny 3}}},
\end{eqnarray}
where $\Theta(x)$ is the step-function. It ensures that ${\huge P}_{1,1}$
drops out of the system (\ref{derivP10})-(\ref{derivP00}) for $V<V_c$. 
The expression for current that accounts for both cases reads
\begin{equation}
\label{current}
I=\frac{e{\huge P}_{0,1}}
{\tau_{\mbox{\tiny 3}}}+
\frac{e
{\huge P}_{1,1}}
{\tau_{\mbox{\tiny 3}}}\Theta(V-V_c).
\end{equation}
In the stationary regime, the master equations can be solved for 
$ {\huge P}_{n_1,n_2}$ in terms of 
$ {\huge P}_{0,1}$, namely, 
$ {\huge P}_{0,0}=
{\huge P}_{0,1}\tau_{\mbox{\tiny 1}}/
\tau_{\mbox{\tiny 3}}$, $ {\huge P}_{1,0}= {\huge P}_{0,1}\left[
\tau_{\mbox{\tiny 2}}/
\tau_{\mbox{\tiny 3}}+\left(
\tau_{\mbox{\tiny 2}}/
\tau_{\mbox{\tiny 1}}\right)\Theta(V-V_c)\right]$,
$ {\huge P}_{1,1}=
{\huge P}_{0,1}\left(\tau_{\mbox{\tiny 3}}/
\tau_{\mbox{\tiny 1}}\right)\Theta(V-V_c)$. 
 With the normalization condition, 
$ {\huge P}_{0,0}+ {\huge P}_{0,1}+ {\huge P}_{1,0}+ {\huge P}_{1,1}\Theta~(V-~V_c)=1$,
we get
\begin{equation}
\label{P01tauI_1}
{\huge P}_{0,1}=
\frac{\tau_{\mbox{\tiny 3}}}
{
\tau_{\mbox{\tiny 1}}+
\tau_{\mbox{\tiny 2}}+
\tau_{\mbox{\tiny 3}}+
\frac{\tau_{\mbox{\tiny 3}}}{\tau_{\mbox{\tiny 1}}}
\left(\tau_{\mbox{\tiny 2}}+\tau_{\mbox{\tiny 3}}\right)}
\end{equation}
for $V>V_c$ and
\begin{equation}
\label{P01tauI_2}
{\huge P}_{0,1}=
\frac{\tau_{\mbox{\tiny 3}}}
{
\tau_{\mbox{\tiny 1}}+
\tau_{\mbox{\tiny 2}}+
\tau_{\mbox{\tiny 3}}
}
\end{equation}
for $V<V_c$. Substituting Eqs. (\ref{P01tauI_1}), (\ref{P01tauI_2})
into Eq.  (\ref{current}) yields\cite{raikh92,kinkhabwala00} 
\begin{equation}
\label{exact1}
I_2=\frac{e~(\tau_{\mbox{\tiny 1}}+\tau_{\mbox{\tiny 3}})}
{ (\tau_{\mbox{\tiny 1}}+\tau_{\mbox{\tiny 3}})^2-
\tau_{\mbox{\tiny 1}}\tau_{\mbox{\tiny 3}}+
\tau_{\mbox{\tiny 2}}(\tau_{\mbox{\tiny 1}}+\tau_{\mbox{\tiny 3}})}, ~~~V>V_c,
\end{equation}
and 
\begin{equation}
\label{exact2}
I_1=\frac{e}{\tau_{\mbox{\tiny 1}}+\tau_{\mbox{\tiny 2}}+
\tau_{\mbox{\tiny 3}}},~~~~V<V_c.
\end{equation}
To find $I_c$, we note that in the regime
of current dominated by two-electron 
transitions only ${\huge P}_{0,1}$ and ${\huge P}_{1,0}$
are nonzero. They are related via the master equation 
\begin{equation}
\label{derivP10tauc}
\frac{d\!~{\huge P}_{1,0}}{d\!~t}=
-\frac{
{\huge P}_{1,0}}
{\tau_{\mbox{\tiny 2}}}+
\frac{
{\huge P}_{0,1}}
{\tau_{\mbox{\tiny c}}}
\end{equation}
and the normalization condition  ${\huge P}_{0,1}+ {\huge P}_{1,0}=1$. 
In the stationary regime, these two relations yield for  $I_c={\huge P}_{1,0}/
\tau_{\mbox{\tiny 2}}$ the following expression
\begin{equation}
\label{exact3}
I_c=\frac{e}{\tau_{\mbox{\tiny 2}}+\tau_{\mbox{\tiny c}}}.
\end{equation}

The $I-V$ characteristics of a chain contains three steps, see Fig. 4.
Their magnitudes, $I_c/I_d$, $I_1/I_c$, and $I_2/I_1$, depend on the spatial 
positions of the sites.
From Eqs. (\ref{exact1})-(\ref{exact3}) it is easy to
analyze the dependence of these magnitudes on $d_1$ and $d_2$.
First, we
note that the product 
$\tau_{\mbox{\tiny 1}}\tau_{\mbox{\tiny 2}}\tau_{\mbox{\tiny 3}}\propto \exp(2D/a)$
does not depend on $d_1, d_2$. Secondly,  the product 
$\tau_{\mbox{\tiny 1}}\tau_{\mbox{\tiny 3}}$ contains the same tunneling exponent as
$\tau_{\mbox{\tiny c}}$.
These observations allow one to express the magnitudes of the steps as follows
\begin{equation}
\label{ratio1}
\frac{I_c}{I_d}=\frac{\exp\Bigl(\frac{D}{a}\Bigr)}
{2\cosh\Bigl\{\frac{2}{a}\Bigl[\frac{D}{2}-\bigl(d_1+d_2\bigr)\Bigr]\Bigr\}},
\end{equation}
\begin{equation}
\label{ratio2}
\frac{I_1}{I_c}=\frac
{1+\exp\biggl\{\frac{4}{a}\Bigl[\frac{D}{2}-\bigl(d_1+d_2\bigr)\Bigr]\biggr\}}
{1+2\cosh\Bigl(\frac{d_1-d_2}{a}\Bigr)\exp\biggl\{\frac{3}{a}\Bigl
[\frac{D}{3}-\bigl(d_1+d_2\bigr)\Bigr]\biggr\}},
\end{equation}
\vspace{3mm}
\end{multicols}
\widetext
\vspace*{-0.2truein} \noindent \hrulefill \hspace*{3.6truein}
\begin{equation}
\label{ratio3}
\frac{I_2}{I_1}= \frac
{2\cosh\Bigl(\frac{d_1-d_2}{a}\Bigr)\Biggl[2\cosh\Bigl(\frac{d_1-d_2}{a}\Bigr)+
\exp\biggl\{\frac{3}{a}\Bigl[\bigl(d_1+d_2\bigr)-\frac{D}{3}\Bigr]\biggr\}\Biggr]}
{\Bigl[2\cosh\Bigl(\frac{d_1-d_2}{a}\Bigr)\Bigr]^2+
2\cosh\Bigl(\frac{d_1-d_2}{a}\Bigr)\exp\biggl\{\frac{3}{a}\Bigl[
\bigl(d_1+d_2\bigr)-\frac{D}{3}\Bigr]\biggr\}-1}.
\end{equation}
\hspace*{3.6truein}\noindent \hrulefill 
\begin{multicols}{2}
\noindent
From Eqs. (\ref{ratio1})-(\ref{ratio3}) it is easy to see
that the magnitude of the first step is large, as long 
as $d_1,d_2 \gg a$. Concerning the second and the third steps,
the more pronounced they are, the smaller the difference 
$\vert d_1-d_2\vert$. Even for a completely symmetric arrangement, 
$d_1=d_2=d$, the second step is present only if $d<D/4$. Precisely at $d=D/4$,
we have $\tau_{\mbox{\tiny 2}}=\tau_{\mbox{\tiny c}}$.
As $d$ decreases, the magnitude of the second step grows
first as $\exp{\Bigl\{8\bigl(D/4-d\bigr)/a\Bigr\}}$ for $D/4 >d > D/6$,
and then slower, as $\exp{\Bigl\{2\bigl(D/2-d\bigr)/a\Bigr\}}$,
for $d<D/6$. The origin of this growth is that for smaller $d$ 
the waiting time, $\tau_{\mbox{\tiny c}}$,
for two-electron transitions 
becomes progressively longer than $\tau_{\mbox{\tiny 1}}$, $\tau_{\mbox{\tiny 3}}$.
It is also possible to conclude from Eq.~(\ref{ratio3}) 
that the magnitude of the third step, $I_2/I_1$, does not exceed $4/3$. 
The maximum magnitude corresponds to $d_1=d_2< D/6$.

\subsection{CORRELATION PROPERTIES}

As  mentioned in the Introduction, the advantage of the two-site
model being exactly solvable is that it allows one to analyze the applicability
of the mean-field approach, one of the main ingredients of the theory
of hopping transport. The exact solution captures correlations in the
occupation numbers of sites, whereas within the mean-field description
these correlations are neglected. 
For the ohmic transport, the effect of correlations on the average current
was studied in Ref. \onlinecite{levin82}. 
Below we examine the correlation properties of current
for various regimes of non-ohmic transport.

The average occupation numbers of the sites,  
$\langle n_1 \rangle$ and $\langle n_2 \rangle$,
can be expressed in terms of 
${\huge P}_{n_1,n_2}$ as $\langle n_1 \rangle={\huge P}_{1,0}+{\huge P}_{1,1}$
and $\langle n_2 \rangle={\huge P}_{0,1}+{\huge P}_{1,1}$, respectively.
Neglecting correlations is equivalent to setting zero the difference
\begin{eqnarray}
\label{definition}
& &\langle n_1 n_2 \rangle -\langle n_1\rangle \langle n_2\rangle=
{\huge P}_{11}-\Bigl({\huge P}_{1,0}+{\huge P}_{1,1}\Bigr)\Bigl({\huge P}_{0,1}+{\huge P}_{1,1}\Bigr)
\nonumber \\
&=&{\huge P}_{0,0} {\huge P}_{1,1}-{\huge P}_{0,1}{\huge P}_{1,0}.
\end{eqnarray}
We will characterize correlations by the parameter
$\kappa~=~\left[\langle n_1 n_2 \rangle/\langle n_1\rangle \langle n_2\rangle\right]~-~1$,
which, using Eq. (\ref{definition}), can be presented as
\begin{equation}
\label{kap}
\kappa=\frac{{\huge P}_{0,0} {\huge P}_{1,1}-{\huge P}_{0,1}{\huge P}_{1,0}}
{\langle n_1\rangle\langle n_2\rangle},
\end{equation}
so that for uncorrelated transport $\kappa=0$.
It can now be shown that, for $V>V_c$, 
the {\em exact} equations (\ref{derivP10})-(\ref{derivP00})
can be cast in the following form
\begin{eqnarray}
\label{rate_kappa}
\frac{d\!~\langle n_1\rangle}{d\!~t}&=&\frac{
\bigl(1-\langle n_1\rangle\bigr)}{\tau_{\mbox{\tiny 1}}}-
\frac{\langle n_1\rangle
\bigl(1-\langle n_2\rangle\bigr)}{\tau_{\mbox{\tiny 2}}}+
\kappa\frac{\langle n_1\rangle\langle n_2\rangle}
{\tau_{\mbox{\tiny 2}}},\cr
\frac{d\!~\langle n_2\rangle}{d\!~t}&=&\frac{\langle n_1\rangle
\bigl(1-\langle n_2\rangle\bigr)}{\tau_{\mbox{\tiny 2}}}-
\frac{\langle n_2\rangle}{\tau_{\mbox{\tiny 3}}}-
\kappa\frac{\langle n_1\rangle\langle n_2\rangle}
{\tau_{\mbox{\tiny 2}}}.
\end{eqnarray}
In this form, the mean-field description emerges upon neglecting
the last terms in the right-hand sides. Solving the resulting system
of equations for average occupations,
we reproduce the result of Ref.~\onlinecite{glazman88}
\begin{equation}
\label{MF1}
I_2^{\mbox{\tiny MF}}\!=\! 
\frac{e~\langle n_2\rangle}{\tau_{\mbox{\tiny 3}}}\!=\!
\frac{2e}{\tau_{\mbox{\tiny 1}}+
\tau_{\mbox{\tiny 2}}+\tau_{\mbox{\tiny 3}} +
\left[\left(\tau_{\mbox{\tiny 1}}+\tau_{\mbox{\tiny 2}}+
\tau_{\mbox{\tiny 3}}\right)^2-4\tau_{\mbox{\tiny 1}}
\tau_{\mbox{\tiny 3}}\right]^{1/2}}.
\end{equation}
From Eqs. (\ref{exact1}) and (\ref{MF1}) it can be seen that
the ratio $I_2/I_2^{\mbox{\tiny MF}}$ can be expressed 
as a function of a dimensionless parameter
$z=\tau_{\mbox{\tiny 2}}\left(\tau_{\mbox{\tiny 1}}^{-1}+
\tau_{\mbox{\tiny 3}}^{-1}\right)$ as follows
\begin{equation}
\label{Ratio}
F(z)\!=\!\frac{I_2}{I_2^{\mbox{\tiny MF}}}=\frac{b+z+\sqrt{(b+z)^2-4b}}
{2(b+z-1)},~~b=\frac{(\tau_{\mbox{\tiny 1}}+\tau_{\mbox{\tiny 3}})^2}
{\tau_{\mbox{\tiny 1}}\tau_{\mbox{\tiny 3}}}.
\end{equation}
The function $F(z)$ is plotted in Fig. 5 for different values of the
ratio $\tau_{\mbox{\tiny 1}}/\tau_{\mbox{\tiny 3}}$. We see that,
unlike the ohmic case\cite{levin82}, the exact current can exceed the mean-field value
if $\tau_{\mbox{\tiny 2}}$ is sufficiently long. Moreover, for $z=1$,
we have $I_2=I_2^{\mbox{\tiny MF}}$ for {\em arbitrary} ratio   
$\tau_{\mbox{\tiny 1}}/\tau_{\mbox{\tiny 3}}$.

\begin{figure}
\narrowtext
{\epsfxsize=8.5cm
\epsfysize=7.0cm
\centerline{\epsfbox{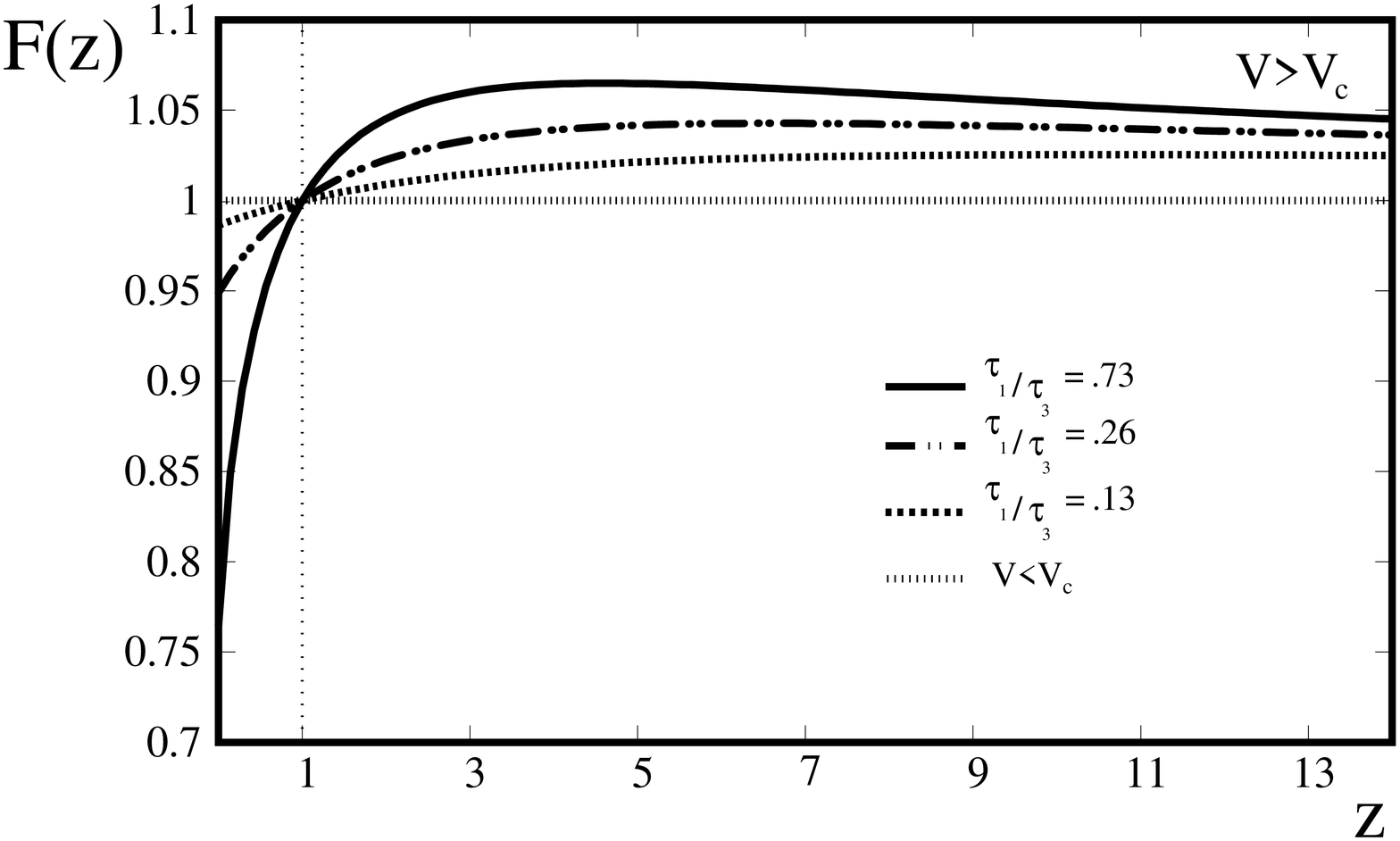}}}
\protect\vspace{0.5cm}
\caption{Ratio of exact and mean-field currents is plotted 
versus a dimensionless parameter
$z=\tau_2\left(\tau_1^{-1}+\tau_3^{-1}\right)$ for  
$\tau_1/\tau_3=0.73$ (solid line);
$\tau_1/\tau_3=0.26$ (dashed-dotted line); and
$\tau_1/\tau_3=0.13$ (dashed line).}
\label{fig3}
\end{figure}
The explanation of these facts can be obtained 
if we express the correlation parameter, $\kappa$, 
in terms of $\tau_{\mbox{\tiny 1}}$, $\tau_{\mbox{\tiny 2}}$,
and $\tau_{\mbox{\tiny 3}}$, using Eq. (\ref{P01tauI_1})
\begin{equation}
\label{correlator1_dimless}
\kappa= \frac{\tau_{\mbox{\tiny 1}}[\tau_{\mbox{\tiny 1}}
\tau_{\mbox{\tiny 3}}-\tau_{\mbox{\tiny 2}}
(\tau_{\mbox{\tiny 1}}+\tau_{\mbox{\tiny 3}})]}
{(\tau_{\mbox{\tiny 1}}+\tau_{\mbox{\tiny 3}})
(\tau_{\mbox{\tiny 1}}\tau_{\mbox{\tiny 2}}
+\tau_{\mbox{\tiny 2}}\tau_{\mbox{\tiny 3}}+
\tau_{\mbox{\tiny 3}}^2)}=\frac{\tau_{\mbox{\tiny 1}}^2
(1-z)}{(\tau_{\mbox{\tiny 1}}+\tau_{\mbox{\tiny 3}})
(\tau_{\mbox{\tiny 1}}z+\tau_{\mbox{\tiny 3}})}.
\end{equation}
We see that for $z=1$ we have $\kappa=0$, so that the current 
is effectively uncorrelated.  For this reason we have 
$I_2=I_2^{\mbox{\tiny MF}}$ when  $z=1$. For $z<1$, the correlation
parameter is positive, resulting in $I_2<I_2^{\mbox{\tiny MF}}$.
Similarly,  $I_2>I_2^{\mbox{\tiny MF}}$ for $z>1$ is the manifestation
of the fact that the current is {\em negatively} correlated.
The latter has a simple explanation: for long $\tau_{\mbox{\tiny 2}}$
(i.e. for $z\gg 1$), as an electron tunnels onto site $1$ from the
left lead, site $2$ is likely to be {\em empty}. Then the electron
does not have to ``wait'' extra time beyond  $\tau_{\mbox{\tiny 2}}$
to proceed to site $2$. The mean-field description does not capture
this effect, thus causing $I_2^{\mbox{\tiny MF}}<I_2$.

For $V<V_c$ the mean-field description reduces to the following equations
for the average occupation numbers  
\begin{eqnarray}
\label{rate1}
 \frac{d\!~\langle n_1\rangle}{d\!~t}&=&
\frac{\bigl(1-\langle n_1\rangle\bigr)\bigl[1- \langle n_2\rangle
\bigr]}
{\tau_{\mbox{\tiny 1}}} \cr
 &-&  \frac{\langle n_1\rangle
\bigl(1-\langle n_2\rangle\bigr)}
{\tau_{\mbox{\tiny 2}}},\cr
\noindent\frac{d\!~\langle n_2\rangle}{d\!~t}&=&\frac{\langle n_1\rangle
\bigl(1-\langle n_2\rangle\bigr)}{\tau_{\mbox{\tiny 2}}}-
\frac{\langle n_2\rangle}{\tau_{\mbox{\tiny 3}}},
\end{eqnarray}
where the factor $(1- \langle n_2\rangle)$ in the first equation
expresses the fact that $l\rightarrow 1$ is possible only when
site $2$ is empty. It easy to see that these equations
yield 
$I_1^{\mbox{\tiny MF}}=e/(\tau_{\mbox{\tiny 1}}+\tau_{\mbox{\tiny 2}}+
\tau_{\mbox{\tiny 3}})$, i.e. the {\em exact} value of current in
the regime $V<V_c$, see Eq. (\ref{exact2}). The explanation why
neglecting the correlations does not change the average current
can be inferred from rewriting Eqs. (\ref{rate1}) in the following form
\begin{eqnarray}
\label{rate2}
 \frac{\tau_{\mbox{\tiny 1}}}{\tau_{\mbox{\tiny 1}}+\tau_{\mbox{\tiny 2}}}         
\frac{d\!~\langle n_1\rangle}{d\!~t}&=&\Bigl[
\frac{\bigl(1-\langle n_2\rangle\bigr)}
{\tau_{\mbox{\tiny 1}} +\tau_{\mbox{\tiny 2}}}
-  \frac{\langle n_1\rangle}{\tau_{\mbox{\tiny 2}}}\Bigr]+
\frac{\langle n_1\rangle\langle n_2\rangle}{\tau_{\mbox{\tiny 2}}},
\cr
 \frac{d\!~\langle n_2\rangle}{d\!~t}&=&\Bigl[
-\frac{\langle n_2\rangle}
{\tau_{\mbox{\tiny 3}}}
+ \frac{\langle n_1\rangle}{\tau_{\mbox{\tiny 2}}}\Bigr]
-\frac{\langle n_1\rangle\langle n_2\rangle}{\tau_{\mbox{\tiny 2}}}.
\end{eqnarray}
Note now, that if one neglects the product $\langle n_1\rangle\langle n_2\rangle$
in the right-hand sides, then the mean-field equations (\ref{rate2})
reduce to the {\em exact} equations (\ref{derivP10})-(\ref{derivP00}) for $V<V_c$,
since for $V<V_c$ we have $\langle n_1\rangle={\huge P}_{1,0}$ and
$\langle n_2\rangle={\huge P}_{0,1}$.
Then the fact that both Eqs. (\ref{rate2}) and (\ref{derivP10})-(\ref{derivP00})
yield the same value of $\langle n_2\rangle$, and thus the same current $I_1$,
is a consequence of the observation that, upon adding  the two equations
 (\ref{rate2}), the products  $\langle n_1\rangle\langle n_2\rangle$ {\em cancel 
each other}.

As follows from  Eq. (\ref{definition}), in two other regimes
with $I=I_1$ and $I=I_c$ we have $\kappa =-1$. This reflects the fact that the
passage of current occurs via a {\em single} repeating cycle in both regimes.
In particular, the cycle for $I=I_c$ consists of two-electron transition
$(l,2) \rightarrow (1,r)$ followed by a single-electron transition $1\rightarrow 2$.
The corresponding waiting times obey the Poisson distributions with 
averages $\tau_{\mbox{\tiny c}}$ and $\tau_{\mbox{\tiny 2}}$, 
respectively. This allows one to find the Fano factor 
of the current noise in the regime $I=I_c$  to be 
$(\tau_{\mbox{\tiny c}}^2+\tau_{\mbox{\tiny 2}}^2)/
(\tau_{\mbox{\tiny c}}+\tau_{\mbox{\tiny 2}})^2 < 1$.
Current noise in the Hubbard-correlated regime, $I=I_2$,
was studied in detail in Ref. \onlinecite{kinkhabwala00}.

\section{OHMIC REGIME}

Similarly to sequential hopping, the transport 
regime dominated by two-electron transitions
is activationless only if applied bias, $V$, is
high enough. Indeed, as seen from Fig. 3, the
domain where the current is equal to $I_c$
exists only for finite $V$.
As the bias is reduced, the current assumes
activational character in both regimes. 
Then the question arises as to whether the
two-electron regime survives in the ohmic
limit $V \ll T$, where $T$ is the temperature.
We will address this question in the following
sequence. First we derive an expression, analogous to 
Eq. (\ref{exact3}), for
the ohmic resistance due to two-electron 
transitions. Then we will demonstrate that
this expression applies to the two-site chain
within a certain domain of energies, 
$\varepsilon_1, \varepsilon_2$, and positions
$d_1, d_2$ of the sites.
 
In order to generalize  the derivation of Eq. (\ref{exact3})
to the ohmic case, we introduce, along with the 
time $\tau_{\mbox{\tiny 2}}$, the time of a ``reverse'' hop
\begin{equation}
\label{ohmicTau2}
\tau_{\mbox{\tiny 2}}^{-}=\tau_{\mbox{\tiny 2}}\exp\left\{\frac{\varepsilon_1-
\varepsilon_2}{T}\right\}.
\end{equation}
We also note that in the regime $V \ll T$ the forward two-electron
transition, illustrated with forward arrows in Fig. 2, acquires
an activation energy, $\varepsilon_1-\varepsilon_2 -V$,
whereas the reverse two-electron transition $(1,r)\rightarrow (l,2)$
becomes {\em activationless} with the characteristic time,
$\tau_{\mbox{\tiny c}}$, calculated as above. Therefore, the time of the 
forward two-electron transition is 
\begin{equation}
\label{ohmicTauc}
\tau_{\mbox{\tiny c}}^{+}=
\tau_{\mbox{\tiny c}}\exp\left\{\frac{\varepsilon_1-
\varepsilon_2-V}{T}\right\}, 
\end{equation}
where $\tau_{\mbox{\tiny c}}$ is now given by
\begin{equation}
\label{evaluate1}
\frac{1}{\tau_{\mbox{\tiny c}}}=\frac{\Gamma_l\Gamma_r}{\pi\hbar~U}
\Biggl[\mbox{\large$\Phi$}\Bigl(\varepsilon_2/U\Bigr)-
\mbox{\large$\Phi$}\Bigl(\varepsilon_1/U\Bigr)\Biggr],
\end{equation}
with $\Phi(x)$ defined by Eq. (\ref{phi}).
The latter expression was derived assuming that 
$\left(\varepsilon_1-\varepsilon_2\right)\gg T$.
As we will see below, in the ohmic regime, the most interesting 
situation corresponds to $\left(\varepsilon_1-\varepsilon_2\right)\sim  T \ll U$.
Then one has to perform the integration in the general expression Eq. (\ref{form})
[with $\varepsilon_1$ and $\varepsilon_2$ interchanged due to the redefinition of 
$\tau_{\mbox{\tiny c}}$ by  Eq. (\ref{ohmicTauc})]
using the explicit form of the distribution functions in the leads. This yields 
\begin{equation}
\label{eqn36}
\frac{1}{\tau_{\mbox{\tiny c}}}\!=\!
\frac{\Gamma_l\Gamma_r U^2}{2\pi\hbar\varepsilon_1^2\left(\varepsilon_1+U\right)^2}\cdot
\frac{\varepsilon_1-\varepsilon_2}{1-\exp{\Bigl[\left(\varepsilon_2-\varepsilon_1\right)/T\Bigr]}}.
\end{equation}

As in the previous section, we assume here that the sites are in
the ratchet configuration, i.e. their energies are within the
interval $\bigl[0,~U\bigr]$ below the Fermi level, so that interaction-elevated
energies of both sites are above the Fermi level.
In the regime dominated by two-electron transitions, the only sets
of the occupation numbers involved in the passage of current are $(0,1)$ and $(1,0)$.  
In the presence of reverse transitions, the master equation
Eq.~(\ref{derivP10tauc}) assumes the form
\begin{equation}
\label{derivP01ohmic}
\frac{d\!~{\huge P}_{1,0}}{d\!~t}=
-{\huge P}_{1,0}\Biggl(\frac{1}{
\tau_{\mbox{\tiny c}}}+
\frac{1}{\tau_{\mbox{\tiny 2}}}\Biggr)+
{\huge P}_{0,1}\Biggl(\frac{1}{
\tau_{\mbox{\tiny c}}^{+}}+
\frac{1}{\tau_{\mbox{\tiny 2}}^{-}}\Biggr).
\end{equation}
With the normalization condition,  ${\huge P}_{0,1}+ {\huge P}_{1,0}=1$, this
yields the following stationary solution for the probability ${\huge P}_{0,1}$
\begin{equation}
\label{ohmicP01}
{\huge P}_{0,1}=\frac{
\tau_{\mbox{\tiny c}}^{-1}+
\tau_{\mbox{\tiny 2}}^{-1}}
{\tau_{\mbox{\tiny c}}^{-1}+
\tau_{\mbox{\tiny 2}}^{-1}+
\left(\tau_{\mbox{\tiny c}}^{\mbox{\tiny +}}\right)^{-1}+
\left(\tau_{\mbox{\tiny 2}}^{\mbox{\small -}}\right)^{-1}}.
\end{equation}
In the presence of reverse hops, the stationary current 
can be calculated, {\em e.g.}, as the difference of the
forward and reverse currents between sites $1$ and $2$, namely,  
$I_c={\huge P}_{1,0}/\tau_{\mbox{\tiny 2}}-{\huge P}_{0,1}/\tau_{\mbox{\tiny 2}}^{-}$.
With the use of Eq. (\ref{ohmicP01}), we then obtain
\begin{equation}
\label{ohmicIc}
I_c=\frac{\left(\tau_{\mbox{\tiny c}}^{\mbox{\tiny +}}\right)^{-1}
-\tau_{\mbox{\tiny c}}^{-1}
\left(\tau_{\mbox{\tiny 2}}/\tau_{\mbox{\tiny 2}}^{-}\right)}        
{1+\left(\tau_{\mbox{\tiny 2}}/
\tau_{\mbox{\tiny c}}\right)+
\left(\tau_{\mbox{\tiny 2}}/\tau_{\mbox{\tiny c}}^{+}\right)+
\left(\tau_{\mbox{\tiny 2}}/
\tau_{\mbox{\tiny 2}}^{-}\right)}.
\end{equation}
In the ohmic regime, $V\ll T$, the numerator of Eq. (\ref{ohmicIc})
is proportional to $V$. Substituting  (\ref{ohmicTauc}) and (\ref{ohmicTau2})
into (\ref{ohmicIc}), we find the ohmic resistance caused by two-electron hops
\begin{equation}
\label{resistance}
R_c^{-1}(T)=\frac{I_c}{V}=\frac{e}{T\bigl(\tau_{\mbox{\tiny 2}}+
\tau_{\mbox{\tiny c}}\bigr)}\Bigl[\exp{\Bigl(\frac{\vert\varepsilon_1-
\varepsilon_2\vert}{T}\Bigr)}+1\Bigr]^{-1}.
\end{equation}

Although in our derivation we assumed that 
$\varepsilon_1>\varepsilon_2$, by introducing  $|...|$
into  Eq. (\ref{resistance}) and also into Eq. (\ref{eqn36}), 
we ensure that it applies
to $\varepsilon_1<\varepsilon_2$ as well.
Comparing this result to ``non-ohmic'' Eq. (\ref{exact3}),
we conclude that basically in the ohmic regime the current, $I_c$, 
acquires the activation energy, $\vert \varepsilon_1-\varepsilon_2\vert$. 
We would like to emphasize that the necessary condition for 
both Eq. (\ref{exact3}) and Eq. (\ref{resistance}) to apply
is that  sites $1$ and $2$ are in the ratchet configuration.

While in the non-ohmic regime with $l\rightarrow 1$ and 
$2\rightarrow~r$ blocked, the two-electron hops 
constituted  the {\em only} channel for activationless current flow, 
it is not obvious whether the ``two-electron resistance'' Eq. (\ref{resistance})
can determine the net resistance in the ohmic regime. 
We address this question below.

Note first that, whether the current is governed by sequential hopping
or by two-electron hops, the transfer of electrons between the leads
necessarily includes the transition $1\rightarrow 2$ (or $2\rightarrow 1$).
Thus, the ``competition'' between two-electron and sequential hopping is
decided by the dominant mechanism of reoccupation of  sites $1$ and $2$
{\em from the leads} during intervals {\em between} the hops  
$1\rightarrow 2$ (or $2\rightarrow 1$). 
If this dominant mechanism is two-electron transitions $(l,2) \rightarrow (1,r)$
or reverse, then the longest waiting time for this transition, $\tau_{\mbox{\tiny c}}^{\mbox{\tiny +}}$,
must be shorter than {\em all} characteristic single-electron times 
involved in the reoccupation of the sites. For  single-electron transitions
$l\longleftrightarrow 1$ the shortest such time is given by
\begin{equation}
\label{pm1}
\ln~\tau_{\mbox{\tiny 1}}^{\pm}=\frac{2}{a}\left(\frac{D}{2}-d_1\right)+\frac{\Delta_1}{T},
\end{equation}
where $\Delta_1=$ min $\left\{\vert\varepsilon_1\vert,\vert\varepsilon_1+U\vert\right\}$ 
is the smallest activation energy of the hops $1\rightarrow l$  (when site $2$ is empty) and
 $l\rightarrow 1$ (when site $2$ is occupied).
Analogously, for transitions $r\longleftrightarrow 2$, the shortest reoccupation time is
defined as
\begin{equation}
\label{pm3}
\ln~\tau_{\mbox{\tiny 3}}^{\pm}=\frac{2}{a}\left(\frac{D}{2}-d_2\right)+\frac{\Delta_2}{T},
\end{equation}
with $\Delta_2=$ min $\left\{\vert\varepsilon_2\vert,\vert\varepsilon_2+U\vert\right\}$.
The times (\ref{pm1}) and (\ref{pm3}) should be compared to the longest time of
the two-electron transition, which in terms of $d_1$ and $d_2$ can be expressed as follows
\begin{equation}
\ln~\tau_{\mbox{\tiny c}}^{\mbox{\tiny +}}= \frac{2}{a}\Bigl(D-d_1-d_2\Bigr) +
\frac{\vert\varepsilon_1-\varepsilon_2\vert}{T}.
\end{equation}
Then the conditions $\tau_{\mbox{\tiny c}}^{\mbox{\tiny +}}<\tau_{\mbox{\tiny 1}}^{\pm}$
and $\tau_{\mbox{\tiny c}}^{\mbox{\tiny +}}<\tau_{\mbox{\tiny 3}}^{\pm}$ can be cast in the form
\begin{eqnarray}
\label{ohmictriangle}
d_2>\frac{D}{2}-\frac{a \Bigl(\Delta_1-\vert \varepsilon_1-\varepsilon_2\vert\Bigr)}{2T},\cr
d_1>\frac{D}{2}-\frac{a \Bigl(\Delta_2-\vert \varepsilon_1-\varepsilon_2\vert\Bigr)}{2T}.
\end{eqnarray}
It is seen from Eq. (\ref{ohmictriangle}) that the most favorable situation
for two-electron transport to dominate is when the energies of the sites are
close to each other, i.e.  when $\vert \varepsilon_1-\varepsilon_2\vert\ 
\ll \vert \varepsilon_1\vert,\vert \varepsilon_2\vert$. In this situation
we have $\Delta_1\approx \Delta_2 =\Delta$, so that the conditions 
(\ref{ohmictriangle}) can be presented as $d_1,d_2>\Bigl[D/2-a\Delta/2T\Bigr]$.
Graphically these conditions are represented by the vertical and horizontal
lines on the plane $\{d_1, d_2\}$, see Fig. 6.
\begin{figure}[!t]
\narrowtext
{\epsfxsize=8.5cm
\centerline{\epsfbox{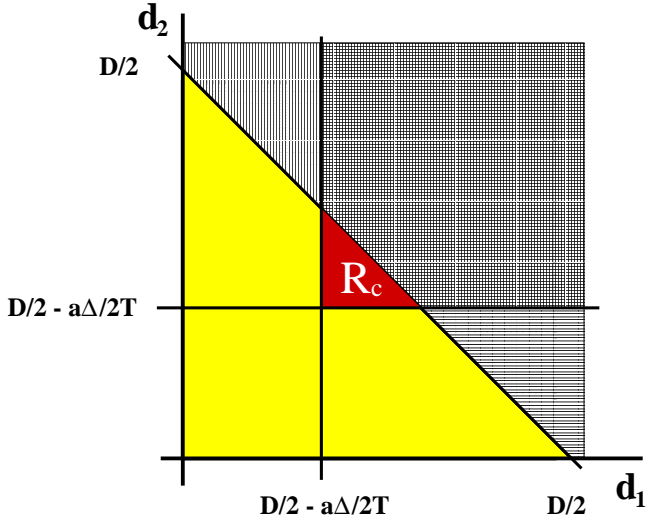}}}
\protect\vspace{0.5cm}
\caption{(Color online)  Triangular region defined by conditions 
$d_1,d_2>D/2-a\Delta/2T$ and $d_1+d_2<D/2$ indicates the
 site positions where two-electron hops 
dominate the transport.} 
\label{fig6}
\end{figure}
\noindent In the domain where these 
conditions are met, Eq. (\ref{resistance}) applies. Note, however,
that when $\tau_{\mbox{\tiny 2}}$ is much longer than $\tau_{\mbox{\tiny c}}$,
it is the hop $1\longleftrightarrow 2$ that determines the resistance
regardless of whether the reoccupation of sites $1$ and $2$ occurs
via two-electron or single-electron transitions.  Thus, the ``true''
two-electron-hopping regime is realized when 
$\tau_{\mbox{\tiny 2}}<\tau_{\mbox{\tiny c}}$. The latter condition
can be presented as $d_1+d_2 < D/2$. It corresponds to the diagonal 
line in Fig. 6. This line together with the horizontal and vertical lines
define a non-empty domain of site positions $d_1$ and $d_2$ if
the condition 
\begin{equation}
\label{last}
D<\frac{2a\Delta}{T} 
\end{equation}
is met. The condition  (\ref{last}) has a transparent physical meaning.
Single-electron hops require shorter tunneling distances, but
involve activation, whereas two-electron hops require tunneling
over longer distances, but they are practically activationless 
when $\varepsilon_1\approx\varepsilon_2$. Thus, in accordance with
Eq. (\ref{last}), the lower the temperature, 
the wider is the range of the barrier thicknesses where 
two-electron hops dominate the transport.

According to the definition of $\Delta$ as the minimal
of the activation energies of the hop onto (or out of) 
Coulomb-shifted and unshifted sites, its value cannot
exceed $U/2$. Then  the most favorable
situation for two-electron-dominated hops is 
$\varepsilon_1 \approx \varepsilon_2 \approx -U/2$.
In this situation,
Eq. (\ref{last}) can be presented as   $D< aU/T$.

\section{CONCLUDING REMARKS}

In the present paper we studied the
effect of interactions on inelastic 
transport through a two-site
chain. In particular, we have identified 
the domain of energies and positions of sites
where the transport occurs via
two-electron transitions, which
are impossible without interactions.
In the ohmic regime, the most  favorable
situation  for the two-electron 
transport regime realizes  
when the energies of the sites 
are close, $\varepsilon_1\approx\varepsilon_2$. Note, however, that
throughout the paper we assumed that
the difference between the site energies,
as well as their energy distance to the
Fermi level, exceed the site widths,
$\Gamma_l$, $\Gamma_r$. This ensures that there is no
resonant tunneling of a single
electron via the two-site configuration\cite{larkin87}.
While {\em interaction-induced} 
resonant tunneling with $\Gamma_l$, $\Gamma_r$
determined by {\em two-electron} virtual 
transitions\cite{kuznetsov96} does not require 
complete alignment of site energies, it does 
 require more than two sites.

For a thick barrier with localized states the 
current is governed by impurity chains
containing many sites\cite{pollak73,tartakovskii87}.
Their energies are close to the Fermi level,
and they are almost equidistant. We do not
expect interactions to have a strong effect
on hopping through these ``optimal'' chains
for the following reason. While the waiting
times for all  hops constituting the chain  
are approximately the same, the net resistance
is determined by the longest of these times.
This is reminiscent of the two-site chain, 
considered above, with $\tau_{\mbox{\tiny 2}}$
longer than $\tau_{\mbox{\tiny 1}}$,  $\tau_{\mbox{\tiny 3}}$,
and $\tau_{\mbox{\tiny c}}$.
As we saw before, under this condition, the fashion 
(correlated or non-correlated) in which
electrons were delivered to/from 
the sites of the main hop was not important.

Unlike inelastic cotunneling, the regime of 
{\em elastic} cotunneling through
a quantum dot\cite{review} does not have an 
analog within the two-site model. This is because
in the case of a quantum dot an electron can be
transferred between the left and right leads only
via the dot. By contrast, in the two-site model, 
an electron can tunnel directly between the leads; 
the amplitude of this process exceeds the amplitude 
of elastic cotunneling via each of the  two sites.  

Note also, that the two-site model 
is too simplistic to capture another
interaction effect on hopping transport,
which is  the effect
of distant ``fluctuators'' away from the
current-carrying path on the 
passage of current\cite{kozub00}.

There exists certain analogy between the regime of transport
by two-electron hops via impurities considered here and
transport by cotunneling through granular 
arrays\cite{feigelman05,beloborodov05,fogler06}. 
The important difference is, however, that in Refs. 
\onlinecite{feigelman05,beloborodov05,fogler06}  
each granule is assumed to contain a large number
of levels; cotunneling  proceeds via these
levels, and it is the finite charging energy of an 
{\em individual} granule that allows this process. 
Interaction {\em between the granules} is essentially
irrelevant in calculating the rates of hops \cite{feigelman05,beloborodov05,fogler06}. 

Since we consider {\em sites} rather than granules,
it is the on-site Hubbard interaction, $U_0$, that 
plays the role of charging energy. This interaction
is assumed infinite in our consideration.  
On the other hand, the {\em intersite} interaction
plays a central role for transport.
As a consequence of the fact that two-electron transitions
in a two-site model are due to the intersite interaction,
the net rate of a hop 
is, essentially, the {\em sum} of the rates for
two-electron and subsequent single-electron transitions.  
By contrast, in the case of granules,
the probability of a particular hop between
two distant granules is proportional to the
{\em product} of  tunneling probabilities  
between {\em all} adjacent granules along the path.



\section{acknowledgements}
The authors are indebted to M. M. Fogler for numerous 
valuable suggestions, which improved the presentation. 
This work was supported by NSF under grant NER~057952.

\end{multicols}
\end{document}